\documentstyle[preprint,aps]{revtex}
\draft
\begin{document}
\title{Adiabatic dynamics of superconducting quantum point contacts}
\author{D. Averin and A. Bardas}
\address{Department of Physics, SUNY Stony Brook, NY 11794}
\maketitle

\begin{abstract}
Starting from the quasiclassical equations for non-equilibrium
Green's functions we derive a simple kinetic equation that governs
ac Josephson effect in a superconducting quantum point contact at
small bias voltages. In contrast to existing approaches the
kinetic equation is valid for voltages with arbitrary time
dependence. We use this equation to calculate frequency-dependent
linear conductance, and dc $I\!-\!V$ characteristics with and
without microwave radiation for resistively shunted quantum point
contacts. A novel feature of the $I\!-\!V$ characteristics is the
excess current $2I_c/\pi$ appearing at small voltages. An important
by-product of our derivation is the analytical proof that the
microscopic expression for the current coincides at arbitrary
voltages with the expression that follows from the Bogolyubov-de
Gennes equations, if one uses appropriate amplitudes of Andreev
reflection which contain information about microscopic structure
of the superconductors.

\end{abstract}

\pacs{PACS numbers: 74.50.+r, 74.80.Fp, 73.20 Dx }
\narrowtext

Point contacts between normal metals have simple Ohmic $I\!-\!V$
characteristics regardless of their electron transparency $D$.
In contrast to this, $I\!-\!V$ characteristics of the
superconducting point contacts may be highly nonlinear even in the
simplest situation of short constriction between two ideal BCS
superconductors, and exhibit a non-trivial dependence on $D$. The
origin of this complexity is the oscillating Josephson current,
which makes electron motion in the contact essentially
inelastic \cite{b1,b2}. Recently, there has been a considerable
progress in calculation of both dc and ac \cite{b3,b4,b5}
components of current in such contacts. However, the results
were limited to the situation when the contact is biased with a
constant (in time) voltage supplied by an ideal source with
vanishing impedance.

It is of interest to generalize the theory of electron transport in
superconducting point contacts to the case of finite impedance of the
voltage source as well as to time-dependent voltages. This
generalization is particularly important in view of the fact that
most experimental realizations of superconducting quantum point
contacts \cite{b6,b7} are based on the superconductor/semiconductor
heterojunctions which typically have relatively large impedance.
Below we develop such a generalization which is valid for small bias
voltages, $V\ll \Delta/e$, where $\Delta$ is the superconductor energy
gap in the electrodes.

We consider a ballistic quantum point contact with characteristic
dimensions much smaller than both the elastic scattering
length and coherence length of the superconducting electrodes.
DC supercurrent in such a contact is known to be carried by the two
discrete energy states with energies $\epsilon_{\pm} =\pm \Delta
\cos \varphi/2$ inside the energy gap \cite{b8,b9}, where $\varphi$
is the Josephson phase difference across the contact. These states
are spatially localized in the contact region because of the
Andreev reflection. At low voltages $V\ll \Delta/e$, dynamics of
these states is slow on the frequency scale given by the energy
gap, $\dot{\varphi} = 2eV/\hbar \ll \Delta/\hbar$, and one could
expect the ac Josephson effect in this regime to be described
in terms of the same two quasistationary states. However, in
contrast to the stationary regime ($V\equiv 0$) when the occupation
of these states is given simply by the equilibrium Fermi-Dirac
probabilities, in the non-stationary situation the occupation
of these states is quite non-trivial.

We first discuss our final result, the kinetic equation which
governs the evolution of occupation probabilities $p_{\pm}$ of the
levels $\epsilon_{\pm}$. (The systematic development leading to this
equation is presented in the last part of the paper.) Because of the
normalization condition $\sum_{\pm} p_{\pm}=1$, it is convenient
to write the kinetic equation in terms of the difference of the two
probabilities, $p(\varphi(t))\equiv p_- -p_+$. Kinetic
equation for $p(\varphi)$ is:
\begin{equation}
\dot{p}(\varphi(t))=\gamma(\epsilon)[n(\epsilon)-p(\varphi(t))]\, ,
\label{1}  \end{equation}
where $n(\epsilon) =\tanh (\epsilon/2T) $ is the equilibrium value
of $p(\epsilon(\varphi))$; $\gamma(\epsilon)$ is the rate
of quasiparticle exchange between the bulk electrodes and discreet
levels in the constriction, and $\epsilon =\epsilon (\varphi) \equiv
\Delta \cos \varphi/2$. The rate $\gamma$ is roughly proportional to
the subgap density of states in the superconducting electrodes;
it vanishes in the ideal BCS case; if the gap is slightly
smeared by finite electron-phonon interaction, $\gamma$ is given
by the following expression \cite{b10,b11}:
\begin{equation}
\gamma(\epsilon)= \alpha \int d\epsilon' \frac{\Theta (\epsilon'^2
-\Delta^2)}{\sqrt{\epsilon'^2 -\Delta^2}}
\frac{(\epsilon-\epsilon')^3 \cosh(\epsilon/2T)}{
\sinh ((\epsilon-\epsilon')/2T) \cosh(\epsilon'/2T)} \, .
\label{2}  \end{equation}
Here $\alpha$ is a constant determined by the parameters of
electron-phonon interaction.
To the kinetic equation (\ref{1}) we should add a ``boundary
condition'' which states that the level occupation reaches
equilibrium as soon as the levels hit the gap edges, $\epsilon =
\pm \Delta$ (Fig.\ 1), that is:
\begin{equation}
p(\varphi )= (-1)^m n(\Delta)  \, ,
\;\;\;\;\; \mbox{for} \; \varphi=2\pi m  \, , \; m=0,\pm 1, ...
\label{3} \end{equation}

We can take into account small reflection coefficient $\cal{R}$ of
the point contact, $\cal{R}$$\ll 1$, by including in the kinetic
equation the Zener transitions between the two levels which occur
at the point $\varphi=\pi \mbox{mod}(2\pi)$ with the probability
$\lambda$ \cite{b5}. For vanishing external resistance the transition
probability is $\lambda =\exp \{ -2\pi$$\cal{R}$$\Delta/\hbar \mid
\dot{\varphi} \mid\}$, where $\dot{\varphi}$ is taken at the
transition point. Account of the Zener transitions is achieved by
imposing one more boundary condition on $p(\varphi)$:
\begin{equation}
p(\varphi +0\mbox{sign} (\dot{\varphi}))= (2\lambda-1)
p(\varphi -0\mbox{sign} (\dot{\varphi})) \, ,
\;\;\;\;\; \mbox{for} \; \varphi =\pi \, \mbox{mod} (2 \pi) \, .
\label{4} \end{equation}

The function $p(\varphi(t))$ given by the eqs.\ (\ref{1}) -
(\ref{4}) determines the current $I(t)$ in the point contact:
\begin{equation}
I(t)= \frac{\pi \Delta}{e R_N} \sin \frac{\varphi (t)}{2}
p(\varphi(t)) \, ,
\label{5} \end{equation}
where $R_N=\pi \hbar/Ne^2$ is the normal-state resistance of the
contact, and $N$ is the number of transverse modes which are all
assumed to be identical.

Equations (\ref{1}) -- (\ref{5}) allow us to describe dynamics
of the point contact under arbitrary bias conditions. As a first
example, we consider the {\em linear response} of the voltage-biased
point contact to small oscillations of the Josephson phase
difference around some stationary point $\varphi_0$, i.e.
$\varphi(t)=\varphi_0 + \varphi_{\omega} e^{-i\omega t}$, $\mid \!
\varphi_{\omega} \! \mid \ll 1$. Equations (\ref{1}), (\ref{5})
with this $\varphi(t)$ give that the current oscillates around the
stationary value \cite{b12}:
\begin{equation}
I_s (\varphi_0) =\frac{\pi \Delta}{eR_N} \sin \frac{\varphi_0}{2}
\tanh (\frac{\epsilon_0}{2T} ) \, ,
\label{56} \end{equation}
so that $I=I_s + I_{\omega}e^{-i\omega t}$, and the frequency
dependent linear conductance is:
\begin{equation}
Y(\omega)=\frac{I_{\omega}}{V_{\omega}} = \frac{2eiI_{\omega}}{\hbar
\omega \varphi_{\omega}} = \frac{2\pi \Delta}{\hbar \omega R_N}
[\frac{\Delta}{4T} \frac{ \sin^2 (\varphi_0/2)}{ \cosh^2
(\epsilon_0/2T) } \frac{\gamma(\epsilon_0)}{\omega +
i\gamma(\epsilon_0) }+ \frac{i}{2} \cos \frac{\varphi_0}{2}
\tanh ( \frac{\epsilon_0}{2T} )  ] \, .
\label{6} \end{equation}
where $\epsilon_0 = \epsilon(\varphi_0)$. This equation generalizes
the corresponding expression obtained by Zaitsev \cite{b1} for
large temperatures $T\gg \Delta$. (Note that eq.\ (\ref{6}), as well
as the kinetic equation (\ref{1}) give only the leading terms in
small relaxation rate $\gamma$.)

We also can define the linear response to small dc voltage, $V\ll
\hbar \gamma/e$. In this case the phase increases indefinitely,
$\varphi(t)=\varphi_0 + 2eVt/\hbar$, but deviation of the occupation
probability $p$ from equilibrium is still small. For such an
evolution of $\varphi$ eqs.\ (\ref{1}) and (\ref{5}) give:
\begin{equation}
I(t) =I_s(\varphi(t))+ \frac{V}{R_N}
\frac{\pi \Delta^2}{2\hbar \gamma (\epsilon) T}
\frac{\sin^2 (\varphi( t)/2)}{\cosh^2 (\epsilon /2T)}   \, .
\label{7} \end{equation}
where $\epsilon=\epsilon(\varphi(t))$. This equation is the
generalization of the recent result \cite{b13} to energy-dependent
relaxation rate $\gamma(\epsilon)$. We see that both types of
linear response are sensitive functions of $\gamma (\epsilon)$ and
therefore they may be used to measure the subgap density of states
in the superconductors.

Non-linear response of the ballistic ($\cal{R}$$=0$) point contact
to constant voltage $V$ at zero temperature can also be obtained
directly from eqs.\ (\ref{1}), (\ref{5}). At $T=0$ we obtain:
\begin{equation}
I(\varphi(t)) =I_c \mbox{sign} V \sin \frac{\varphi}{2}
\left\{ \begin{array}{ll}
1\, , \;\;\;\; & 0< \varphi <\pi \, , \\ 2\exp (-\int_{t_0}^t
\gamma(\epsilon)dt)-1 \, , \;\;\;\; & \pi < \varphi <2\pi \, ,
\end{array} \right.
\label{78} \end{equation}
where $t_0$ is the time at which $\varphi (t)=\pi$. Equation
(\ref{78}) shows that the effect of finite relaxation rate $\gamma$
is very similar to the effect of finite reflection probability
$\cal{R}$ (see eq.\ (11) in \cite{b5}).

As another application of the kinetic equation (\ref{1}) we consider
{\em resistively shunted} superconducting quantum point contact
biased by an external current $I_e$ (see inset in Fig.\ 3). For such
a bias condition, the evolution equation for $\varphi$ reads:
\begin{equation}
\frac{\hbar\dot{\varphi}(t)}{2eR_e} = I_e-I(\varphi(t)) \, ,
\label{8} \end{equation}
where current $I(\varphi(t))$ should be calculated from eqs.\
(\ref{1}) - (\ref{5}) self-consistently with $\varphi(t)$, and $R_e$
is the shunting Ohmic resistance, which adiabatic approximation
requires to be small: $R_e \ll R_N$. We limit ourselves to low
temperatures, $T\ll \Delta$.

For very small external resistances $R_e/R_N\ll \{ (\hbar \gamma/
\Delta)\,,\cal{R}\}$, the rate of $\varphi$ evolution is small and
the current $I(\varphi)$ is given by the stationary
relation (\ref{56}). Applying this relation in eq.\ (\ref{8}),
we conclude that the dc $I\!-\!V$
characteristic of the point contact is given by the same relation
as in the standard RSJ model (see, e.g., \cite{b14}, Sec.\ 4.2):
\begin{equation}
I=(I_c^2+(\frac{V}{R_e})^2)^{1/2} - \frac{V}{R_e} \, .
\label{9} \end{equation}
In the opposite limit of relatively large external resistances $R_e/
R_N \gg \{ (\hbar \gamma/\Delta)\, ,\cal{R} \}$, the current-phase
relation $I(\varphi)$ is the same as with no relaxation \cite{b5}:
\begin{equation}
I(\varphi) =I_c \mbox{sign}(V) \mid \sin \frac{\varphi}{2} \mid \, ,
\label{10} \end{equation}
and the $I-V$ characteristic is given by the following
relations:
\begin{equation}
V=\frac{(I_e^2 -I_c^2)^{1/2}}{R_e} \frac{\pi}{4\arctan \sqrt{
(I_e+I_c)/(I_e-I_c)]} } \, ,\;\;\;\;   I=I_e-V/R_e\, .
\label{11} \end{equation}
The main qualitative
difference between expression (\ref{11}) and the quasistationary
expression (\ref{9}) is the excess current: $I\rightarrow 2/\pi
I_c$ for $V\gg I_cR_e$ in eq.\ (\ref{11}), whereas in eq.\
(\ref{9}) the current obviously vanishes at large voltages. Note
that eqs.\ (\ref{9}) and (\ref{11}), and Figs.\ 2 and 3 below
give the $I-V$ characteristics in the form (i.e.,
without the linear term $V/R_e$) that is directly applicable to
typical bias conditions of point contacts which are not shunted
intentionally. Under these conditions there is no shunting
resistance at zero frequency, but there is finite impedance of
the biasing leads in series with the contact at frequencies of the
Josephson oscillations. It is known that this situation can be
reduced to the RSJ model by simple subtraction of the dc current
through the resistor (see, e.g., \cite{b14}, Sec.\ 12.4).

Figure 2 shows how $I-V$ characteristics evolve from
eq.\ (\ref{9}) into eq.\ (\ref{11}) with increasing external
resistance. The curves were calculated numerically from eq.\
(\ref{78}) and (\ref{8}) assuming no reflection in the
point contact ($\cal{R}$$=0$), and also assuming that
$\gamma$ is a phenomenological constant independent of energy.
In the case when this transition is driven not by finite
relaxation rate $\gamma$ but by finite reflection $\cal{R}$
the curves look qualitatively very similar.

Figure 3 shows dc $I-V$ characteristics of the point contact
under microwave irradiation, $I_e(t)=I_0+A\cos(\Omega t)$,
which exhibit the usual Shapiro steps at voltages
$V_{k,m} = (k/m)\hbar \Omega/2e$. In the standard way (see
Ref.\ \cite{b14}, Sec.\ 10.2) one can get that at large
frequencies the $I-V$ characteristic in the vicinity of the steps
has a hyperbolic shape (similar to eq.\ (\ref{9})) with the step
height $I_c \mid a_m J_k(2eR_eAm/\hbar \Omega) \mid$. Here $J_k$
are the Bessel functions, and $a_m$ are the Fourier expansion
coefficients of $I(\varphi)$,
\[ a_m=\frac{8}{\pi}\frac{(-1)^{m} m}{1-4 m^{2}} \, , \;\;\;\;\;
\mbox{or} \;\;\;\; a_m=\frac{4}{\pi} \frac{1}{4 m^{2}-1}\, , \]
for the quasistationary current (\ref{56}), and for the current
given by eq.\ (\ref{10}), respectively. From these expressions we
can conclude that the height of the subharmonic steps ($m\neq 1$)
which are the hallmark of the presence of higher harmonics in
$I(\varphi)$ depends strongly on the value of external resistance.
It increases for small external resistance due to the current
discontinuity at $\varphi=\pi\, \mbox{mod}\, 2\pi$.

Now we briefly outline the major steps leading to our basic kinetic
equation (\ref{1}). We start with the quasiclassical equation
for non-equilibrium Green's functions
of the superconductors (for general introduction to this
technique see, e.g., \cite{b15}). The Green's functions can be
represented as $G^{(0)}+G$, where $G$ is a space-dependent
non-equilibrium addition to the equilibrium part $G^{(0)}$ that
is constant inside each electrode. For short constrictions,
equations for the retarded and advanced parts of $G$ read
\cite{b10,b1}:
\begin{equation}
iv_F \frac{\partial G_{R,A} }{\partial z}=[H_{R,A},G_{R,A}]  \, ,
\;\;\;\; H_{R,A}=(\delta_{R,A}+i\gamma_{el}) G^{(0)}_{R,A} \, ,
\label{12} \end{equation}
where $\delta_{R,A} \equiv ((\epsilon \pm i\gamma_1)^2-(\Delta
\pm i \gamma_2)^2)^{1/2}$; $\gamma_{1,2}$ and $\gamma_{el}$ are,
respectively, inelastic and elastic scattering rates, $v_F$ is the
Fermi velocity, and coordinate $z$ measures the distance from the
point contact ($z=0$) into the electrodes ($z\rightarrow \pm
\infty$). All functions in eq.\ (\ref{12}) are matrices in the
electron-hole space; for instance, $G^{(0)}_{R,A} (\epsilon,
\epsilon') =[((\epsilon \pm i\gamma_1)\sigma_z +(\Delta \pm i
\gamma_2)i\sigma_y)/\delta_{R,A} ]  \delta(\epsilon-\epsilon')$,
with $\sigma$'s here and below denoting Pauli matrices.

The functions $G$ should decay inside the electrodes (at $z
\rightarrow \infty$). If we perform ``rotation'' in the
electron-hole space diagonalizing $G^{(0)}_{R,A}$:
\begin{equation}
G^{(0)}_{R,A} (\epsilon, \epsilon') \rightarrow U_{R,A} (\epsilon)
G^{(0)}_{R,A} (\epsilon, \epsilon') U_{R,A}^{-1} (\epsilon') =
\pm \sigma_z \delta(\epsilon-\epsilon') \, , \;\;\;\;
U_{R,A} = \frac{1+a_{R,A}\sigma_x}{\sqrt{1-a_{R,A}^{2}}} \, ,
\label{13} \end{equation}
where $a_{R,A}\equiv (\epsilon \pm i \gamma_1 - \delta_{R,A})/
(\Delta \pm i \gamma_2)$, eq.\ (\ref{12}) shows then explicitly that
solutions decaying inside the electrodes should have the following
matrix form:
\begin{equation}
G_R^{(1,2)} = U_R^{-1} u_R^{(1,2)} \sigma_{\pm} U_R\, , \;\;\;\;
G_A^{(1,2)} = U_A^{-1} u_A^{(1,2)} \sigma_{\mp} U_A\, ,
\label{14} \end{equation}
where $\sigma_{\pm} = \sigma_x\pm i\sigma_y$, and $G^{(1,2)}$
denote the function in the first ($z<0$) and the second ($z>0$)
electrode, respectively.

The total Green's functions should be continuous at the point
contact ($z=0$). Imposing this condition and taking into account
that there is a voltage drop $V$ between the two electrodes of the
point contact we can determine the functions $u_{R,A}^{(1,2)}$ in
eq.\ (\ref{14}). At small voltages, $V\rightarrow 0$, we get then for
the total Green's functions $\bar{G}_{R,A}=G_{R,A}+G^{(0)}_{R,A}$
at $z=0$:
\begin{equation}
\bar{G}_{R,A} (\epsilon,t) = \int \frac{d \epsilon'}{2\pi}
\bar{G}_{R,A} (\epsilon+ \frac{\epsilon'}{2},\epsilon -
\frac{\epsilon'}{2}) e^{i\epsilon' t} =
\frac{i}{2\pi}\frac{\sigma_z \cos[\varphi/2 - \arccos
(\epsilon /\Delta )] +i\sigma _y}{\sin [\varphi/2 -\arccos
(\epsilon /\Delta) \pm i0]} \, ,
\label{15} \end{equation}
where $\bar{G}_{R,A}$ depend on time $t$ via the time
dependence of the Josephson phase difference $\varphi$, $\varphi=
2eVt/\hbar +\varphi_0$ . In eq.\ (\ref{15}) we neglected
the relaxation rates $\gamma_{1,2}$ assuming that they are small.
This is a legitimate approximation since, as usual, the effect of
small energy relaxation on the occupation probabilities (i.e.,
on $G_K$) is much more important than the effect on the density of
states. For $\bar{G}_{R,A}$ given by eq.\ (\ref{15}), the latter
effect would be a small broadening of the Andreev-bound level.

One can check directly from eq.\ (\ref{15}) that this equation agrees
with the stationary Green's functions calculated first by Kulik and
Omel'yanchuk \cite{b12}. In particular, in the subgap range
$\mid \!\epsilon\! \mid < \Delta$ it corresponds precisely to one
of the two Andreev-bound discrete energy levels: $\mbox{Re}
\bar{G}_{R,A} \propto \delta (\epsilon-\Delta\, \mbox{sign} (\sin
\varphi /2) \cos \varphi/2 )$.  Since the evolution equation
(\ref{12}) and, consequently, eq.\ (\ref{15}) refer to electrons
moving in the positive $z$-direction ($p_z>0$) this is the level that
carries current in one direction. Evolution equation for electrons
with $p_z<0$ differs only by the sign in front of $v_F$. In this case
we get that $\bar{G}_{R,A}$ corresponds to the energy level at
$\epsilon=-\Delta\, \mbox{sign} (\sin \varphi /2)\cos \varphi/2 $.

To find the current $I(t)$ in the point contact we need to calculate
the Keldysh component $G_K$ of the Green's function \cite{b10,b1}:
\begin{equation}
I(t)= \frac{\pi }{2R_N} \int d\epsilon
\mbox{Sp} \{ \sigma_z (\bar{G}_K^{(p_z>0)}(\epsilon,t) -
\bar{G}_K^{(p_z<0)}(\epsilon,t) \} \, .
\label{16} \end{equation}
Equation for $G_K$ is:
\begin{equation}
iv_F\frac{\partial G_K}{\partial z}=H_RG_K+H_KG_A-G_RH_K-G_KH_A\, ,
\;\;\;\; H_K=H_Rn-nH_A \, ,
\label{17} \end{equation}
where $H_{R,A}$ are defined in eq.\ (\ref{12}), and $n$ is the
equilibrium quasiparticle distribution, $n(\epsilon,\epsilon')=
\tanh (\epsilon /2T) \delta(\epsilon-\epsilon')$.
This equation shows that $G_K$ can be written as \cite{b3}
$G_K=G_Rn-nG_A+G_H$, where $G_H$ is the part that satisfies the
homogeneous equation:
\begin{equation}
iv_F\frac{\partial G_H}{\partial z}=H_RG_H-G_HH_A\, ,
\label{18} \end{equation}
Following the same steps that led to eq.\ (\ref{14}) we get that
$G_H$ should have the following matrix form:
\begin{equation}
G_H^{(1,2)} = U_R^{-1} u_H^{(1,2)} (1 \pm \sigma_z) U_A\, .
\label{19} \end{equation}

Imposing again the continuity condition at $z=0$, we calculate
$u_H^{(1,2)}$ and then find the current at arbitrary voltages:
\[ I(t)=\sum_k I_k e^{i2keVt/\hbar} \, , \]
\begin{equation}
I_k =\frac{1}{eR_N} \left[ eV \delta_{k0} - \int d\epsilon
\tanh( \frac{\epsilon}{2T} ) (1-\mid a_R(\epsilon )\mid^2)
\sum_{n=0}^{\infty} \prod_{m=1}^n \mid a_R(\epsilon +m eV)\mid^2
\prod_{m=n+1}^{n+2k} a_R(\epsilon +m eV)
\right]  \, .
\label{20} \end{equation}

Equation (\ref{20}) has the same form as the corresponding
expression that follows from calculations based on the
Bogolyubov-de Gennes equations \cite{b5}. The only difference is
that the function $a_R(\epsilon)$ which has the meaning of
generalized Andreev reflection amplitude
now contains full information about the microscopic properties
of the superconducting electrodes, and is, in
general, different from its ``ideal'' BCS value. In the particular
case considered here it includes finite energy relaxation rates
$\gamma_{1,2}$. For small $\gamma$'s,
part of the eq.\ (\ref{20}) related to the dc current ($k=0$)
reduces to the so-called BTK expression for the current \cite{b2}.
To the best of our knowledge, this is the first explicit proof that
the widely used BTK approach is equivalent to the microscopic
theory of electron transport in short ballistic constrictions.

Finally, to obtain the kinetic equation (\ref{1}) we consider the
Green's function $\bar{G}_K$ in the limit $V\rightarrow 0$, when
it is given by the following expression:
\[ \bar{G}_K (\epsilon,t)= \frac{\Delta}{2} \mid \sin
\frac{\varphi}{2} \mid \delta (\epsilon- \Delta \, \mbox{sign} (\sin
\varphi /2) \cos \varphi/2) N(\epsilon,t) \, , \]
where
\begin{equation}
N(\epsilon ,t) = n(\epsilon ) + \int^{\Delta}_{\epsilon }d
\epsilon' \frac{\partial n}{\partial \epsilon'}\exp \{ -\int_{
\epsilon }^{\epsilon'}\frac{ d \epsilon'' \hbar \gamma
(\epsilon'')}{eV \sqrt{\Delta^2-\epsilon''^2}} \} \, .
\label{21} \end{equation}
Here $\hbar \gamma \equiv 2(\gamma_1 -(\epsilon /\Delta )
\gamma_2)$, so that $\gamma$ is given by the eq.\ (\ref{2}).

Comparison of this expression for $\bar{G}_K$ with the subgap
density of states that follows from eq.\ (\ref{15}) shows directly
that $N(\epsilon,t)$ has the meaning of quasiparticle distribution,
so that $(1-N(\epsilon,t))/2$
can be interpreted as an occupation probability of one of the two
Andreev-bound levels inside the gap. Equation (\ref{21}) with this
interpretation immediately gives the kinetic eq.\ (\ref{1}) and
the boundary condition (\ref{3}). Indeed, taking into account
the definition of $p$ in the kinetic equation we see that it is
related to $N$ as follows: $p=N \mbox{sign}( \sin \varphi/2)$.
This relation together with the eq.\ (\ref{21}) give the
boundary condition (\ref{3}). Furthermore, differentiating eq.\
(\ref{21}) with respect to energy and making use of the relation
between energy and phase, $\epsilon = \mbox{sign} (\sin \varphi
/2) \Delta \cos \varphi/2$, we get:
\begin{equation}
\frac{2eV}{\hbar} \frac{d N}{d\varphi} = \gamma (\epsilon )
(N(\epsilon)-n(\epsilon)) \, .
\label{22} \end{equation}
If we express $N$ in this equation in terms of $p$ we finally
arrive at eq.\ (\ref{1}). Although we assumed so far that the
voltage $V$ is constant in time, it is obvious that
the evolution equations (\ref{22}), (\ref{1}) in the differential
form are valid for arbitrary time dependence of the voltage, as
long as the voltage itself and the rate of its variations are small.

As a last remark we should mention that thermalization
of the occupation probability $p$ due to the boundary condition
(\ref{3}) is instantaneous only on the long time scale set by the
period of the Josephson oscillation. Crude estimate of the energy
interval $\delta \epsilon$ near the gap edge which determines $p$
is $(\Delta e^2V^2)^{1/3}$, so that the corresponding time scale
of thermalization is $\delta t\simeq \hbar/\delta \epsilon$. In the
relevant limit $eV/\Delta\rightarrow 0$, $\delta t$ is much less
than the period of the Josephson oscillations.

In conclusion, we developed an adiabatic theory of the ac Josephson
effect in short constrictions between two superconductors. The
theory is based on the simple kinetic equation for the
non-equilibrium occupation probabilities of the two Andreev-bound
states localized in the constriction. The kinetic equation is
rigorously derived from the microscopic equations for quasiclassical
Green's functions of the constriction, and can be applied to
situations with arbitrary time dependence of the bias voltage.

The authors gratefully acknowledge useful discussions with
K. Likharev and A. Zaitsev. This work was supported by DOD URI
through AFOSR Grant \# F49620-92-J0508.

\figure{Figure 1. } Energies $\epsilon_{\pm}$ of the two
Andreev-bound levels in a short constriction between two
superconductors as functions of the Josephson phase difference
$\varphi$. Solid dots represent ``thermalization points'' where
occupation of the two level always reaches equilibrium. The
diagram illustrates sign convention for the kinetic equation
\protect (\ref{1}) and the boundary condition (\ref{3}).

\figure{Figure 2. } DC $I-V$ characteristics of the resistively
shunted ballistic superconducting quantum point contact for
various ratios of external Ohmic resistance $R_e$ and
relaxation rate $\gamma$ at zero temperature; $\omega_c \equiv
2eI_c R_e/\hbar$.

\figure{Figure 3. } Effect of microwave radiation with the
amplitude $A=2I_c$ and frequency $\Omega=2\omega_c$ on
the DC $I-V$ characteristics of the resistively shunted
superconducting point contact. The contact parameters are the
same as in Fig.\ 2. From top to bottom, the curves correspond to
$\gamma /\omega_c = 0,\, 1,\, \infty$.


\begin{references}
\bibitem{b1} A. Zaitzev, Sov.\ Phys.\ - JETP {\bf 51}, 111 (1980).
\bibitem{b2} T.M. Klapwijk, G.E. Blonder, and M. Tinkham,
Physica B {\bf 109\&110}, 1657 (1982).
\bibitem{b3} U. Gunsenheimer and A.D. Zaikin, Phys.\ Rev.\ B
{\bf 50}, 6317 (1994).
\bibitem{b4} E.N. Bratus, V.S. Shumeiko, and G. Wendin,
Phys.\ Rev.\ Lett. {\bf 74}, 2110 (1995).
\bibitem{b5} D. Averin and A. Bardas, Phys.\ Rev.\ Lett. {\bf 75},
1831 (1995).
\bibitem{b6} K.-M. H. Lenssen, P.C.A. Jeekel, C.J.P.M. Harmans,
J.E. Mooij, M.R. Leys, J.H. Wolter, M.C. Holland, in: {\em Coulomb
and interference effects in small electronic structures}, Ed.\
by D.C. Glattli, M. Sanquer, J. Tr\^an Thanh V\^an (Editions
Frontiers, 1994), p.\ 63.
\bibitem{b7} H. Takayanagi, T. Akazaki, and J. Nitta, {\em
Observation of maximum supercurrent quantization in a
superconducting quantum point contact}, preprint (1995).
\bibitem{b8} A. Furusaki and M. Tsukada, Physica B {\bf 165\&166},
967 (1990).
\bibitem{b9} C.W.J. Beenakker and H. van Houten, Phys.\ Rev.\
Lett. {\bf 66}, 3056 (1991).
\bibitem{b10} A.I. Larkin and Yu.N. Ovchinnikov, Sov.\ Phys.\ -
JETP {\bf 41}, 960 (1976).
\bibitem{b11} S.N. Artemenko, A.F. Volkov, and A.V. Zaitsev,
Sov.\ Phys.\ - JETP {\bf 49}, 924 (1979).
\bibitem{b12} I.O. Kulik and A.N. Omel'yanchuk, Sov.\ J. Low
Temp.\ Phys. {\bf 3}, 459 (1977).
\bibitem{b13} A. Levy Yeyati, A. Martin-Rodero, and
J.C. Cuevas, cond-mat 9505102.
\bibitem{b14} K. Likharev, {\em Dynamics of Josephson Junctions
and Circuits} (Gordon and Breach, NY, 1986).
\bibitem{b15} J. Rammer and H. Smith, Rev.\ Mod.\ Phys. {\bf
58}, 323 (1986).
\end{references}
\end{document}